\documentclass[aps,prb,reprint,superscriptaddress,showpacs,amsmath]{revtex4-1}
\usepackage{graphicx}
\usepackage{color}
\usepackage{dcolumn}

\usepackage[extra]{tipa}
\setcitestyle{square,numbers}

\begin{document}

\title{Comparison of computer-algebra strong-coupling perturbation theory and
dynamical mean-field theory for the Mott-Hubbard insulator in high dimensions}

\author{Martin Paech}\email{martin.paech@itp.uni-hannover.de}
\affiliation{Institut f\"{u}r Theoretische Physik, Leibniz Universit\"{a}t Hannover, 30167 Hannover, Germany} 
\affiliation{Academy of Computer Science, 43-300 Bielsko-Bia{\l}a, Poland}
\author{Walter Apel}
\affiliation{Physikalisch-Technische Bundesanstalt, 38116 Braunschweig, Germany} 
\affiliation{Institut f\"{u}r Theoretische Physik, Leibniz Universit\"{a}t Hannover, 30167 Hannover, Germany}
\author{Eva Kalinowski}
\affiliation{Academy of Computer Science, 43-300 Bielsko-Bia{\l}a, Poland}
\author{Eric Jeckelmann}
\affiliation{Institut f\"{u}r Theoretische Physik, Leibniz Universit\"{a}t Hannover, 30167 Hannover, Germany}

\date{\today}

\begin{abstract}
We present a large-scale combinatorial-diagrammatic computation
of high-order contributions to the strong-coupling Kato-Takahashi perturbation
series for the Hubbard model in high dimensions.
The ground-state energy of the Mott-insulating phase is determined exactly
up to the 15--th order in $1/U$. 
The perturbation expansion is extrapolated to infinite order and the critical
behavior is determined using the Domb-Sykes method. 
We compare the perturbative results with two dynamical mean-field
theory (DMFT) calculations
using a quantum Monte Carlo method and a density-matrix renormalization group method 
as impurity solvers. The comparison demonstrates the excellent agreement and accuracy of  
both extrapolated strong-coupling perturbation theory and
quantum Monte Carlo based DMFT,  even close to the critical coupling where the Mott 
insulator becomes unstable.

\end{abstract}

\pacs{71.10.Fd,71.27.+a,71.30.+h,02.10.Ox}

\maketitle

\section{Introduction}

The Kato-Takahashi strong-coupling perturbation theory (SCPT)~\cite{kato,takahashi} and the dynamical 
mean-field theory (DMFT)~\cite{vol11,vol12} are two powerful methods for studying strongly correlated quantum
many-body systems such as the Mott insulating phase~\cite{mot90,geb97}
 found in the Hubbard model~\cite{hub63,gut63,kan63} with on-site interaction $U$.
In high dimensions the Kato-Takahashi SCPT can be calculated exactly up to high orders
in $1/U$ using a combinatorial-diagrammatic approach~\cite{blu05,kal12} while the DMFT scheme becomes exact in principle.
Early comparisons of both methods~\cite{nis04,blu05} showed a very good agreement deep in the Mott 
insulating phase.
However, they let some open questions about the relative accuracy of different
impurity solvers for DMFT  and the properties
of the Mott insulating phase close to the critical coupling $U_c$ where it becomes unstable.
Here, we report on a large-scale computer-algebra calculation of higher orders in the SCPT series expansion and 
the resolution of this issue.

\subsection{Mott insulator and Hubbard model}

The nature of Mott insulators without long-range magnetic order 
is a long-standing open problem in the theory of strongly correlated quantum systems.
Theoretically, Mott insulating phases can be found in strongly interacting fermion~\cite{mot90,geb97} or 
boson~\cite{fis89} systems as well as in fermion-boson mixtures~\cite{alt12}. 
Experimentally, non-magnetic Mott insulators have been found in a layered organic 
insulator~\cite{kur05}. In this triangular-lattice material, the frustration of the 
antiferromagnetic spin exchange coupling prevents the formation of a long-range magnetic
order. A Mott insulator phase can also be realized in an atomic gas trapped in an 
optical lattice~\cite{gre02,blo08}. Despite decades of extensive research, the properties of 
Mott insulators and, more generally, the transition from a Mott insulator to a metallic 
(fermion system) or superfluid (boson system) phase are only partially understood and thus actively investigated.

The Hubbard model with repulsive on-site interaction~$U\geq 0$ and nearest-neighbor
hopping term~$t \geq 0$ is a basic lattice
model for studying the physics of strongly interacting electrons, in particular the Mott metal-insulator 
transition.~\cite{geb97} At half filling (one electron per lattice site) the ground state is a Mott insulator for 
strong interaction~$U/t$, while it is a Fermi gas in the non-interacting limit~$U=0$. 
Thus, the system must undergo a metal-insulator transition at some coupling
$U^{\text{MIT}}_c \geq 0$.
If the lattice geometry (defined by hopping integrals between lattice sites) causes a strong frustration 
of the effective antiferromagnetic exchange coupling between electron spins, the ground state is paramagnetic. 
Thus, this model can describe the transition from a paramagnetic Mott insulator to a metallic state. 
The Bose-Hubbard model is an extension of the Hubbard model to boson systems, which can be used to describe 
the transition from a Mott insulator to a superfluid.~\cite{blo08}

Here we consider the Hamiltonian for interacting electrons
\begin{eqnarray}
H & = & T + U D \nonumber \\
\label{eq:hamiltonian}
T & = & - \sum_{\langle i, j\rangle}
\sum_{\sigma=\uparrow, \downarrow}
\frac{t_{\sigma}}{\sqrt{Z}}
\left(c^{\dag}_{i, \sigma}c^{\phantom{\dag}}_{j, \sigma}+c^{\dag}_{j,
\sigma}c^{\phantom{\dag}}_{i, \sigma}\right)  \\ 
D & = &  \sum_{i} n_{i, \uparrow} n_{i, \downarrow}
\nonumber
\end{eqnarray}
where $c^{\dag}_{i, \sigma}$ and $c^{\phantom{\dag}}_{i, \sigma}$ are the standard
fermion creation and annihilation operators for an electron with spin $\sigma$
on the site with index $i$ and $n_{i, \sigma} = c^{\dag}_{i, \sigma} c^{\phantom{\dag}}_{i, \sigma}$
are the local density operators.
To reach the limit of infinite dimensions, we choose a Bethe lattice with connectivity~$Z$ in the limit 
$Z\to\infty$. Thus, the last sum in~(\ref{eq:hamiltonian}) runs over the $L$ lattice sites while the first sum runs
over the $LZ/2$ pairs of nearest-neighbor sites. We restrict the present study
to a half-filled system (i.e., one electron per site  on average).
The Hubbard model corresponds to equal hopping terms for both spins
$t_{\uparrow} = t_{\uparrow} = t$ while the Falicov-Kimball model\cite{don92} corresponds to
a single mobile electron species $t_{\uparrow} = t > 0$ and $t_{\downarrow} = 0$.
We will set the unit of energy by~$t\equiv 1$.

\subsection{Dynamical mean-field theory}

DMFT and its generalizations have become a leading approach for studying correlated electronic 
systems.~\cite{vol11,vol12} 
They have been combined with the density functional theory~\cite{vol12,pav11} to
perform first-principles calculations. 
This DFT+DMFT approach is increasingly used to investigate materials with strong electronic correlations 
such as transition metals and their oxides. 
More recently, the application of DMFT to quantum chemistry problems has been explored.~\cite{lin11}

The Mott metal-insulator transition in the Hubbard model
has been extensively investigated with DMFT.~\cite{vol11,bul01,blu02,nis04,gar04,kar05}
On a Bethe lattice with infinite coordination number, DMFT studies have revealed
a first-order quantum phase transition from a Fermi liquid to
a Mott insulator as $U$ increases. The ground state is metallic up to a critical coupling
$U^{\text{MIT}}_c$ (which is close to $5.8$ according to QMC-DMFT calculations~\cite{blu02})
and becomes insulating above this value.
However, the Mott insulating state remains metastable down to a critical
coupling 
$U_c < U^{\text{MIT}}_c$ where the Mott-Hubbard gap closes.
Thus, it still influences the system properties for $U_c < U < U^{\text{MIT}}_c$
and in real systems, it should be observable in experiments such as
time-resolved spectroscopy.

In the DMFT approach, a bulk system is mapped onto an effective self-consistent quantum impurity system. 
In the case of the Hubbard model, this is the well-known single impurity Anderson model (SIAM).
This mapping becomes exact in the limit of infinite dimensions or coordination number. 
However, solving the quantum impurity problem is a very hard task in most cases. 
Various ``impurity solvers'' can be used to compute the SIAM properties numerically.
For instance, numerical renormalization group (NRG)~\cite{bul01},
density matrix renormalization group (DMRG)~\cite{geb03,nis04,gar04,kar05},
and quantum Monte Carlo (QMC)~\cite{blu05,blu02} methods have been used successfully for this purpose.
Thus, in practice, one has to solve the self-consistent impurity problem numerically and recursively. 
This introduces errors which are difficult to estimate within the DMFT scheme. 
Therefore, reliable results obtained with other methods are highly desirable to validate the DMFT approach and 
evaluate its accuracy, even in the limit of infinite dimensions. 
So far, besides DMFT computations, most reliable results for the Hubbard model in the limit of high dimensions 
have been obtained using weak~\cite{mue89,geb03} and strong coupling perturbation 
theory~\cite{nis04,blu05,kal12}.
Additionally, the Kato-Takahashi SCPT can be used to solve the DMFT self-consistency 
equation.~\cite{ruh11}

\subsection{Strong-coupling perturbation theory}

Series expansion methods, especially perturbative approaches, constitute basic
theoretical tools of physics.~\cite{oit06,apel,hag07,che04,yan10,yan11,kla12}
They are often used to investigate strongly correlated lattice models such as the Hubbard model. 
An attractive feature of these methods is that they are often well suited for the use of high-performance 
computer algebra. 
Thus, one can take advantage of the computational power of modern supercomputers without losing the rigor of 
analytical calculations. 
This can be a decisive advantage over most numerical approaches, which usually have to deal with various 
issues brought by finite-precision algorithms and floating-point arithmetic.

In principle, the Kato-Takahashi perturbation expansion provides us with a systematic method for calculating 
the properties of the Mott insulator in powers of~$1/U$. 
In practice, the number of diagrams contributing to $n$-th order increases exponentially with~$n$  
and thus calculations become rapidly too complex. 
Several years ago, a direct manual calculation yielded the ground-state energy of~(\ref{eq:hamiltonian}) up to the 4-th order and 
the local Green's function up to second order in~$1/U$.~\cite{eas03,kal02,kal02b} 
These results agree well with DMRG-based DMFT simulations for~$U\geq 6$.~\cite{nis04} 
Later, an combinatorial-diagrammatic algorithm was developed to calculate a given order exactly 
using computer algebra. This method allowed one of us to compute the ground-state
energy exactly up to the 11-th order in~$1/U$ using moderate computational
resources.~\cite{blu05,kal12} 
The results agree very well down to~$U=4.8$ with DMFT data obtained using a  
QMC method~\cite{blu05} or a DMRG method~\cite{nis04} as impurity solver. 
However, these studies reached different conclusions regarding the critical coupling $U_c$ where the Mott 
insulator becomes unstable, and the related critical exponent $\tau$. 
Combining QMC-DMFT and SCPT results the first study~\cite{blu05} claimed that 
$U_c\approx 4.78$ (in agreement with other DMFT calculations~\cite{gar04,bul01,blu02}) and $\tau \approx 7/2$
while the DMRG-DMFT-based study~\cite{nis04} concluded that $U_c \approx 4.45$ and  $\tau\approx 5/2$.
In addition, a critical coupling $U_c \approx 4.406$ was deduced
from a perturbative calculation of the Mott-Hubbard gap within the DMFT
approach.~\cite{ruh11}

Unfortunately, even the most recent SCPT study (up to 11-th order in~$1/U$)~\cite{kal12} 
was not sufficient to discriminate between the QMC and DMRG data close to $U_c$.
It is well known that a series expansion truncated at any finite order becomes 
increasingly unreliable as one approaches a critical point (i.e., an analytical singularity)
and that taking higher-order into account improves the reliability.
Thus, in this paper we present a large-scale computer-algebra calculation which allows us to obtain 
the ground-state energy up to the 15-th order in~$1/U$ using the combinatorial-diagrammatic approach. 
The comparison of SCPT and DMFT results reveals that QMC-DMFT agrees much
better than DMRG-DMFT with the 15-th order perturbation theory close to the critical regime
(down to $U=4.8$).
In addition, extrapolating the perturbation expansion to infinite order
using the Domb-Sykes method~\cite{dom57,dom61}
allows us to determine critical coupling 
$U_c\approx 4.76$ and critical exponent $\tau \approx 3.45$
very precisely. This critical coupling also agrees with  previous DMFT calculations.~\cite{blu05,blu02,bul01,gar04}
Moreover, we find that the ground-state energies calculated with the 
extrapolated SCPT and the QMC-DMFT agree perfectly
(within $10^{-7}$ of the band width) even
extremely close to the critical coupling, e.g. for $U-U_c \approx 0.04$.

The rest of this paper is structured as follows. The high-performance computer-algebra SCPT
 is described in the next section. Section~\ref{sec:results} presents the comparison of the SCPT
 results with the DMRG-DMFT and QMC-DMFT data as well as the extrapolated perturbation theory.
Finally, the perspective for further development and applications of the combinatorial-diagrammatic
SCPT are discussed in Sec.~\ref{sec:conclusion}.

\section{Computer-algebra SCPT \label{sec:method}}

\subsection{Kato-Takahashi series expansion}

The ground-state energy per lattice site $E$ of the Hamiltonian~(\ref{eq:hamiltonian})
can be written as a series in power of $1/U$
using the Kato-Takahashi strong-coupling perturbation theory.~\cite{kato,takahashi}
At~$U=\infty$ (or equivalently $t=0$) the ground-state energy is $E_{0}=0$
and the corresponding eigenstates have exactly one electron localized on each lattice site. 
As the spin orientation does not change the energy, the ground state is degenerate.
We denote by $\mathcal{H}_{0}$ the corresponding eigenspace. Its dimension is
$2^{L}$ for an $L$-site lattice. 
Let $P$ be the projection operator onto the lowest-energy eigenspace 
of the Hamiltonian~(\ref{eq:hamiltonian}) at finite $U$.
(Obviously, $P$ is the projector onto $\mathcal{H}_0$ for $U=\infty$.)
Using the generic Kato perturbation theory, Takahashi showed for the
Hubbard model in the strong-coupling limit that this operator can 
be written as a power series in $1/U$ 
\begin{equation}
\label{eq:projector}
P = -\sum_{n=0}^{\infty}\frac{1}{U^{n}}
{\sum_{\{k_{r}\}}}S^{k_{1}}TS^{k_{2}}T\cdots TS^{k_{n+1}}
\end{equation}
with integers $k_r \geq 0$ such that $k_{1}+k_{2}+\cdots+k_{n+1}=n$.
At half filling, this power series has a finite convergence radius $U^{-1}_R > 0$
and that depends on the lattice properties.
The operators $S^k$ are defined by  
\begin{equation}
\label{eq:operator}
S^{k} = 
\begin{cases}
   \displaystyle{-P_{0}} & \text{for $k=0$} \\
\displaystyle{\sum_{d=1}^L \frac{1}{(-d)^{k}}P_{d}}  & \text{for $k>0$}
\end{cases}  
\end{equation}
where $P_{d}$ is the projector on the subspace of states
with exactly~$d$ doubly occupied sites.

Let us assume that $\left \vert \phi_{0} \right \rangle$ is a state in $\mathcal{H}_0$ with the
property $P \left \vert \phi_{0} \right \rangle \neq 0$. 
(There must be at least one such state if the series expansion for $P$ has a finite convergence radius.)
Then the ground-state energy for finite $1/U$ is given by 
\begin{equation}
\label{eq:energy}
E=\frac{1}{L}\frac{\left \langle \phi_{0} \vert PHP \vert \phi_{0}\right
\rangle }{\left \langle \phi_{0} \vert P \vert \phi_{0} \right \rangle } .
\end{equation}
Using~(\ref{eq:projector}), we can expand this energy in power of $1/U$
\begin{equation}
\label{eq:series}
E = \sum_{n=1}^{\infty} \frac{a_n}{U^n}  
\end{equation}
for $U_R < U < \infty$.
Although one cannot write closed formula for the coefficients $a_n$ in general,
they are completely defined by the equations~(\ref{eq:projector}),
(\ref{eq:operator}), and~(\ref{eq:energy}) for a given lattice and a given
hopping operator $T$.

\subsection{Combinatorial-diagrammatic approach}

A combinatorial-diagrammatic approach was developed by one of us to evaluate the
coefficients $a_n$ for the Hubbard and Falicov-Kimball models on a Bethe lattice with an infinite coordination number.~\cite{blu05,kal12} 
Here we summarize the key ideas which are necessary to understand our new implementation of this approach
and we refer the reader to the original publications for more details.
Expanding the projector~(\ref{eq:projector}) in the energy~(\ref{eq:energy}), we see that 
the coefficients $a_n$ are given by sums of expectation values of the form
\begin{equation}
\label{eq:process}
\left \langle \phi_{0} \vert TP_{d_1}TP_{d_2}TP_{d_3}\cdots T \vert \phi_{0}
\right \rangle ,
\end{equation}
which are called processes. 
Different sets $\{ k_r \}$ and $\{ d_r \}$ in eqs.~(\ref{eq:projector}), (\ref{eq:operator}),
and~(\ref{eq:process}) correspond to different processes.
The number of different sets $\{ k_r \}$ in the projector~(\ref{eq:projector}) is $\binom{2n}{n}$ for the $n$--th order
and thus increases exponentially as $\frac{4^n}{\sqrt{\pi n}}$ for high orders $n$.
The number of different sets $\{ d_r \}$ scales as $L^{n-1}$.
Thus the number of possible processes increases exponentially fast with the order $n$. 

The state $\left \vert  \phi_{0}\right \rangle$ is an eigenstate of the double occupation operator
$D$ and thus has a precise number of doubly occupied sites (zero at half filling) while the hopping 
operator $T$ can change the number of doubly occupied sites by at most one. 
Consequently, only some particular sets $\{ d_r \}$ can yield a non-zero
expectation value~(\ref{eq:process}).
Then using the definition of the hopping operator $T$ in Eq.~(\ref{eq:hamiltonian})
each process can be evaluated as the sum of simple expectation values
\begin{equation}
\label{eq:sequence}
\left \langle \phi_{0} \left \vert 
c^{\dag}_{i_1,\sigma_1}c^{\phantom{\dag}}_{i_2,\sigma_2}c^{\dag}_{i_3,\sigma_3}c^{\phantom{\dag}}_{i_4,\sigma_4}
\cdots c^{\phantom{\dag}}_{i_{2l},\sigma_{2l}} \right \vert \phi_{0} \right \rangle
\end{equation}
(called sequences) over lattice sites and spin indices. The sequence length is  $1 \leq l \leq n+1$.
The number of different sequences in a process increases exponentially
fast with the order $n$, roughly as $(2ZL)^l$ for a process containing $l$ hopping operators $T$. 

\begin{table*}[tb]
\caption{\label{table1} Number of different non-zero processes $N_p$,
 number of sequences $N_s$ in~(\ref{eq:sum}), series coefficients $a_n$,  number of processors used,
wall time of calculation, and the total amount of required memory for the non-trivial orders $n$ from 1 to 15
using the preprocessing technique
as well as estimates for the next non-trivial order $n=17$ without preprocessing}    
\begin{ruledtabular}
\begin{tabular}{d|D{.}{.}{0}D{.}{.}{0}D{.}{.}{1}D{.}{.}{0}D{.}{.}{0}@{\hspace{-5\tabcolsep}}l@{\hspace{5\tabcolsep}}d}
n & N_p & N_s & a_n & \text{proc.} & \multicolumn{2}{c}{\text{time}}&
\multicolumn{1}{c}{\text{memory}} \vspace*{-7mm}\\    
&&&&&&& \multicolumn{1}{c}{\text{(Gb)}} \\
\hline
1  & 1     & 2  & \displaystyle{ -\frac{1}{2} }    &  & & & \\
3   & 2     & 20   & \displaystyle{  -\frac{1}{2} }&  & &  & \\
5   & 4     & 648 & \displaystyle{  -\frac{19}{8} }&  & & & \\
7   & 14    & 45472& \displaystyle{  -\frac{593}{32} }&  & & & \\
9   & 48    & 5644880& \displaystyle{  -\frac{23877}{128} }& 1 & 6 & sec. & <1 \\
11  & 193   & 1099056000    & \displaystyle{  -\frac{4496245}{2048} }& 1 & 18 &min.& <1 \\
13  & 795   & 310007943616  & \displaystyle{  -\frac{1588528613}{55296} }& 128 & 12 &min.& 36 \\
15  & 3412  & 119777421416192  & \displaystyle{  -\frac{12927125815211}{31850496} } & 128 & 77 &hours& 528 \\
\hline 
17  & 14803 & <804658\cdot10^{11} & & 262144 & 13 &days& 131 \\
\end{tabular}
\end{ruledtabular}
\end{table*}

Therefore, the evaluation of the coefficients $a_n$ in the series~(\ref{eq:series}) is a
hard computational problem.
However, the computational cost can be greatly reduced if one 
identifies without explicit calculation the many processes and sequences which 
vanish or are equivalent. For instance, processes can be gathered into a small number of classes defined by the positions $r$
of the indices $k_r=0$ in the list $\{ k_r \}$. Then the processes in a given class are only compatible with particular sets $\{ k_r \}$ 
and $\{ d_r \}$ as well as a reduced number of sequences because an 
intermediate state 
\begin{equation}
\label{eq:intermediate} \left . \left .
c^{\dag}_{i_{2r+1},\sigma_{2r+1}}c^{\phantom{\dag}}_{i_{2r+2},\sigma_{2r+2}}
c^{\dag}_{i_{2r+3},\sigma_{2r+3}}
\cdots c^{\phantom{\dag}}_{i_{2l},\sigma_{2l}} \right \vert \phi_{0} \right \rangle
\end{equation}
without any doubly occupied site must be
reached every time 
that $k_r=0 [\Leftrightarrow d_r = 0$ in Eq.~(\ref{eq:process})].  The actual number $N_p$ of inequivalent non-zero processes
in our implementation is given in the second column of Table~\ref{table1} for several orders $n$.

The above discussion is valid for any lattice.
On a Bethe lattice with an infinite coordination number $Z$, however,
the evaluation of sequences and processes can be considerably simplified.
First, it was shown~\cite{kal02} that 
the degeneracy of the singlet ground state for $U=\infty$ is not lifted up to the third order  in~$1/U$
(if one excludes long-range spin orders such as anti-ferromagnetism). 
Assuming that this holds for all orders, we can use any singlet state $\left \vert
\phi_{0}\right \rangle \in \mathcal{H}_0$
or, equivalently, an average over an orthonormal basis of the singlet subspace in $\mathcal{H}_0$. 
(This averaging greatly simplifies the evaluation of sequences as we will see below.)
Moreover, an expectation value~(\ref{eq:sequence}) vanishes unless each creation operator
$c^{\dag}_{i,\sigma}$ in the sequence is matched by a corresponding annihilation 
operator $c^{\phantom{\dag}}_{i,\pm\sigma}$. Thus the site set $\{ i_r \}$
in a sequence~(\ref{eq:sequence}) describes one or more closed paths on the lattice.
As the operator $T$ contains hoppings between nearest-neighbor sites only,
each segment of the path connects two nearest-neighbor sites. 
Moreover, as loops are not possible on a Bethe lattice, the path is self-retracing
and any nearest-neighbor bond can only appear an even number of times in a closed path.
Finally, segments occurring more than twice yield contributions of the order of
$1/Z$ or smaller and thus are negligible in the limit of an infinite
coordination number. 

Some other generic properties can be used to further simplify the problem.
As the total energy must be extensive, disconnected paths resulting from the expansion of  numerator and
denominator in the ratio~(\ref{eq:energy}) must compensate each other and 
thus only sequences corresponding to a single connected path of length $l=n+1$ yield non-zero
contributions in the $n$--th order. This property is sometimes called a linked-cluster theorem but
in the present context it is more a physical argument than a mathematical theorem. 
It was only proven exactly up to the fifth order in $1/U$ in Ref.~\onlinecite{kal02}.
A detailed analysis of the linked-cluster theorem within the Kato-Takahashi 
perturbation theory also indicates that it should be valid for the Mott-insulating phase of the half-filled 
Hubbard model.~\cite{kla12,sch14}
Finally, all lattice sites are equivalent as we assume that the lattice is
infinitely large ($L\rightarrow \infty$).

In summary, the only non-zero contributions to processes~(\ref{eq:process}) come 
from sequences~(\ref{eq:sequence}) corresponding to a single, closed, and
self-retracing path of even length $l=n+1$ through $l/2$ different bonds and 
$l/2+1$ different sites.
This has three important consequences. First, 
only processes corresponding to an even power of the hopping operator $T$,
or equivalently an odd power of $1/U$, contribute to the ground-state energy at
half filling.

Second, after reordering the fermion operators, any contributing sequences can be written as a correlation
function
\begin{equation}
\label{eq:correlations}
\left\langle\phi_{0} \left \vert \prod_{r=1}^{l} c^{\dag}_{i_r, \sigma_r}c^{\phantom{\dag}}_{i_r, \sigma'_r}
 \right \vert \phi_{0}\right \rangle 
\end{equation}
between all sites in a path of length $l$.
Computing the average of the energy~(\ref{eq:energy}) over an orthonormal basis of the singlet 
subspace in $\mathcal{H}_0$ reduces to averaging these correlation functions.
In the thermodynamic limit $L\rightarrow \infty$ all spin configurations of a
finite cluster are equiprobable if one averages over all singlet states
of the full system. Thus we can simply compute the mean value of the correlation function 
over both spin states $\sigma = \uparrow, \downarrow$ at each site in the path. 
We find then that the average value of a correlation function depends only
on the path length $l$ (and so on the order $n$) 
and is simply a rational number $2^{-(n+3)/2}$.

Third, the distinct linked clusters occurring in contributing sequences can be represented one to one
by diagrams called Butcher trees.\cite{kal12,but76} 
Generating all these clusters corresponds to generating all
Butcher trees with $l/2+1=(n+3)/2$~nodes.
We have to use colored Butcher trees with four distinct colors corresponding to the four possible states 
(unoccupied, spin-$\uparrow$, spin-$\downarrow$, and doubly occupied) of an electronic site
to represent the initial spin configurations in $\left \vert \phi_{0}\right
\rangle$ and the intermediate states~(\ref{eq:intermediate}) of a sequence.
After constructing all $n$--th order Butcher trees, all possible sequences on
them can be generated starting from the graph root using a recursive electron
hopping procedure, which takes into account the physical restrictions such as the
Pauli principle.
Therefore, the evaluation of the coefficients $a_n$ in the
series~(\ref{eq:series}) is reduced to a (hard) combinatorial-diagrammatic problem.

\subsection{High-performance computer-algebra implementation}

A combinatorial-diagrammatic approach was used to calculate the coefficients $a_n$ exactly up
to $n=11$ for both the Hubbard model and the Falicov-Kimball model using 
a proof-of-concept computer program and moderate computer
resources.~\cite{blu05,kal12}
Based on the original program, one of us (MP) has implemented 
a high-performance computer-algebra program that calculates the coefficients
$a_n$ exactly for a given odd $n$ (as $a_n=0$ for all even $n$ at half filling).
The coefficients are written 
\begin{equation}
\label{eq:sum}
a_n = r \frac{2^{-(n+3)/2}}{n+1} \sum_{g} \sum_{s} S_s 
\left ( \sum_{p} C_p \prod_{j=1}^{n} d_{j}^{-k_j} \right )
\end{equation}
where $r=1$ for the Falicov-Kimball model and $r=2$ for the Hubbard model.
The first sum runs over all $2^{(n-1)/2}$ process classes $g$.
The sum over the index $s$ represents the sum over all
sequences~(\ref{eq:sequence}) which are compatible with the process class $g$. 
The sequences for the Falicov-Kimball model are a subset of those 
for the Hubbard model.
A set of double occupancy number $\{d_r\}$ is associated with each
sequence $s$ and $S_s=\pm 1$ is the overall sign of this sequence from
the fermion commutation relations.
The total number $N_s$ of all sequences in all classes
is given in the third column of Table~\ref{table1}
up to the order $n=15$.
The sum over the index $p$ in~(\ref{eq:sum}) represents the sum over
all inequivalent elementary processes~(\ref{eq:process})
in the class $g$. These processes are the same for the Falicov-Kimball model and the Hubbard model.
A set of exponents $\{k_r\}$ is associated with each elementary process $p$
and $C_p$ gives the number of equivalent elementary processes.
The total number $N_p$ of all processes in all classes is given in the second column of Table~\ref{table1}
up to the order $n=17$.

We see that $N_s$ increases faster than $n!$.
This apparently disagrees with the above analysis which predicts at most an
exponential increase of the number of processes and sequences with $n$. 
However, the exponential behavior is obtained for a finite lattice $L$ and a finite 
coordination number $Z$ and thus does not preclude a factorial behavior 
in the limits $L,Z \rightarrow \infty$.
For orders up to $n=11$ the computation of (\ref{eq:sum}) can be easily carried out on a workstation. 
Our optimized and parallelized implementation of this combinatorial-diagrammatic
algorithm has allowed us to carry out the calculation up to the order $n=15$ using high-performance
supercomputers.

This implementation represents about $3800$~lines of ANSI-C code (comments excluded).
The program uses only integer numerics with a ``global'' denominator
instead of slower rational numbers. 
 Thus the coefficients $a_n$ are obtained as exact rational numbers. 
We have found that standard $64$-bit integers are enough up to the 13-th order
but $128$-bit ones are required for the 15-th order. For the 17-th order,
integers with $256$ bits or more would be necessary.
In practice, we use the GNU Multiple Precision Arithmetic Library (GMP)~\cite{gmp}, which
provides integers of arbitrary length.

To optimize the program we have implemented the necessary algorithms as fast operations on our bit-coded data structures.
These include standard combinatorial algorithms, e.g. for sorting and permuting,~\cite{knuap} 
as well as more specialized ones, e.g. for computing the overall sign from all fermion operator commutations in a sequence.
Thus our implementation is self-contained and does not require any special software library except GMP.
A further optimization of several algorithms
was achieved thanks to an independent graph-theoretical analysis of the 
representation of sequences by colored Butcher trees.~\cite{gru11,loogen} This analysis was carried
out using the functional programming language \textsc{Haskell}, 
which provides a concise high-level mathematical environment for this purpose, 
e.g. native support for graph structures.
In addition, we have used the On-Line Encyclopedia of Integer Sequences (OEIS)~\cite{oeis}
to analyze the various integer sequences which occur in intermediate steps of
the combinatorial-diagrammatic algorithm 
and thus verify some intermediate results. This analysis has
also helped us to improve the overall program efficiency.

In contrast to the proof--of--concept implementation in Ref.~\onlinecite{kal12},
our implementation consists of a single program.  To generate all contributing sequences $s$
in~(\ref{eq:sum}), it iterates in parallel over all initial states 
(i.e., spin configurations in $\left \vert \phi_0 \right \rangle$)
in an outer loop while an inner loop runs over all combinations of nearest-neighbor pairs
using a recursive electron hopping procedure.
A trade-off between CPU time and memory usage can be achieved if one initially calculates once and stores
the possible sequence weights [i.e.,  the sum over the elementary processes $p$ in Eq.~(\ref{eq:sum})]
for all double occupancy sets $\{d_r\}$. Then during the summation over 
the sequences $s$, one uses the stored weight for the set $\{d_r\}$ corresponding
to each sequence. This preprocessing of sequence weights yields a significant
speed-up at the cost of a higher memory requirement. 
For instance, preprocessing reduces the CPU time by a factor~$3.3$ for the 15-th order,
which offsets the higher computational cost of the GMP library compared to fixed-length integers.
 Theoretically, the required memory for the sequence-generating subroutine increases from  $\propto n^{2}$ without preprocessing to  $\propto 4^{n}$ with preprocessing, while the main program needs a constant amount $\propto 2^{n}$.
In addition, this method results in an unfavorable memory scaling in a parallel computation as
the total memory now increases linearly with the number of processors while it remains almost
constant without preprocessing. 
As an example, for the order $n=15$, $4$~Gb for each processor plus $16$~Gb of
shared memory are used with preprocessing 
against only $16$ Gb overall without preprocessing.

Our code has been designed for running efficiently on parallel supercomputers.
For an efficient handling of shared data by the specialized combinatorial functions,
we have limited ourselves to symmetric multiprocessor (SMP) computer architectures so far. 
Due to the ideal data parallelism in our implementation as well as to a fine-tuned load 
balancing, the scaling behavior of the computing time is excellent 
on all tested machines, at least up to $510$ processor cores on a \texttt{SGI Altix~4700}
and up to $256$ processor cores on the much more powerful \texttt{HP Integrity Superdome~2-32s}.
A simple analysis on basis of Amdahl's law gives $99.93 \%$ code
parallelism.~\cite{hag11}

Nevertheless, calculating the sum~(\ref{eq:sum}) for $n>11$ remain computationally demanding and we have 
to carry out large-scale calculations on SMP machines with hundreds of processors to
obtain the 13--th and 15--th orders in $1/U$. The wall time used and the required memory
are shown in Table~\ref{table1} for calculations performed  using the preprocessing method on a
\texttt{HP~9000~J6750} workstation (orders $n=9$ and $11$)
and a \texttt{HP Integrity Superdome~2-16s} server (orders $n=13$ and $15$).
In Table~\ref{table1}, we also show our estimates for the order $n=17$
using a massively parallel processing (MPP) supercomputer such as the \texttt{IBM~BlueGene/Q} with $262144$~processors.
Note that the preprocessing method could not be used on this computer system without modification
because the available memory per processor would be too low.

The validity and performance of our program were also tested on the Falicov-Kimball
model~\cite{don92} using a state-of-the-art SMP supercomputer (\texttt{HP Integrity Superdome~X}).
As a result, we can confirm that there is no contribution to the ground-state
energy (beyond the first order term) up to the 17--th order in $1/U$, 
i.e. 6 orders higher than in a previous work.~\cite{kal12} 
This test also allows us to estimate the computational cost for the 17-th order
in the Hubbard model.
It shows that $7.7$~Tb of memory would be required with preprocessing on the $240$~processor cores 
of the \texttt{HP Integrity Superdome~X} and that the calculation would last more than one year
(while the test for the Falicov-Kimball model only took $54$~hours).
Therefore, the calculation of the next order in the Hubbard model series
expansion does not seem to be possible with current SMP machines. Nevertheless, it appears 
to be technically possible with current MPP supercomputers such
as the \texttt{IBM~BlueGene/Q}, although we could not use the current implementation of the preprocessing and the computational cost, 
about $13$~days, would still be very high in practice.

\section{Results\label{sec:results}}

\subsection{Comparison of SCPT and DMFT}

In Table~\ref{table1}, we present the coefficients $a_n$ of the power series for the ground-state
energy~(\ref{eq:series}) in the half-filled Hubbard model up to the order $n=15$.
(Only coefficients for odd $n$ are listed as they vanish for all even $n$.)
They agree with those obtained in previous works~\cite{blu05,kal12} up to the 11--th order in $1/U$.
Thus our high-performance program allows us to improve the accuracy of the truncated series by four orders
in $1/U$.
In addition, the average double occupancy per site can be calculated up to the 16--th order in $1/U$ using  the relation
\begin{equation}
\label{eq:doubly}
D(U)=\frac{d}{d U}E(U).
\end{equation}

\begin{figure}[tb]
  \includegraphics[width=0.49\textwidth]{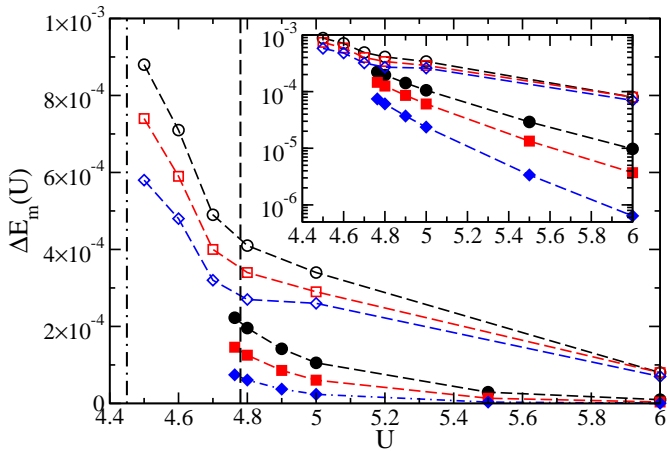}
  \caption{\label{fig1} 
  Absolute differences~(\ref{eq:diff}) between the SCPT and DMFT ground-state
  energies as functions of $U$ for the orders $m=9$ (circle), $m=11$ (square), 
  and $m=15$ (diamond). Open and solid symbols correspond to DMRG-DMFT 
  and QMC-DMFT, respectively. Vertical lines mark the critical coupling $U_c$
  deduced from DMRG-DMFT (dot-dash) and QMC-DMFT (dash) studies. Other lines
  are guides for the eye. The inset shows the same data on a logarithmic scale.
}
\end{figure}

Using the coefficients $a_n$ we can define the partial sums
\begin{equation} 
\label{eq:partial}
E_m(U) = \sum_{n=0}^m \frac{a_n}{U^n} 
\end{equation}
which give the ground-state energy for a given Hubbard interaction $U$ and a
given order of the SCPT up to $m=15$. 
Figure~\ref{fig1} shows the absolute differences 
\begin{equation} 
\label{eq:diff}
\Delta E_m(U)= \left \vert  E_{\text{DMFT}}(U)-E_m(U) \right \vert 
\end{equation}
between the SCPT and DMFT ground-state energies for several $U$ and three
different orders $m$.
We see that $\Delta E_m(U)$ decreases for stronger interaction $U$ and higher order $m$, as expected.
The energy differences are also systematically smaller for QMC-DMFT than for DMRG-DMFT, especially for larger $U$. 
However, this is easily explained by the different precision goal of these two distinct DMFT computations:
DMRG energies~\cite{nis04} were calculated with an accuracy of $10^{-4}$ to $10^{-5}t$
while the QMC data~\cite{blu05} were recorded with an accuracy of $10^{-8}t$.
Moreover, we see in Fig.~\ref{fig1} that the DMRG energy differences become
significantly larger close to the critical value $U_c\approx 4.45$ determined in
a DMRG-DMFT calculation.~\cite{nis04}
This behavior is expected because the SCPT energies
should become rapidly inaccurate as $U$ approaches the convergence radius of the perturbation series.
Surprisingly, the QMC energy differences do not show any sign of a singularity close to the critical
coupling $U_c\approx 4.78$ deduced from QMC-DMFT
computations.~\cite{blu02} 
Therefore, the simple analysis of the energy differences $\Delta E_m(U)$ between
SCPT and DMFT does not allow us to discriminate between both impurity solvers.

In principle, one can examine the convergence of the sequence of partial sums
$\{ E_m(U) ; m=1,3,5, \dots \}$ 
to determine the exact ground-state energy for any given $U > U_c$.
In practice, the extrapolation of a finite number of available terms
$E_m(U)$ to the limit $m \rightarrow \infty$ is often ambiguous.
The case $U=4.8$ is particularly interesting. In Ref.~\onlinecite{blu05} it was shown
that the SCPT ground-state energies $E_m(U)$ up to $m=9$ could be well fitted
with a quadratic function
\begin{equation}
\label{eq:fit}
E_m=E_{\infty} + c_1 x + c_2 x^2
\end{equation}
with 
\begin{equation}
\label{eq:scaling1}
x=\frac{2}{m+1}
\end{equation} 
and the three fit parameters $E_{\infty},c_1,c_2$.
The extrapolated value for $m\rightarrow\infty$ was found
to be $E_{\infty}= - 0.110259$ for $U=4.8$ in excellent agreement with the QMC-DMFT
result $E_{\text{DMFT}}(U=4.8)=  -0.11026919$. However, the choice of the
scaling~(\ref{eq:scaling1}) is rather arbitrary. 
Indeed, in Ref.~\onlinecite{nis04} it was shown that the same SCPT ground-state energies could 
be equally well fitted by a quadratic function~(\ref{eq:fit})
with 
\begin{equation}
\label{eq:scaling2}
x=\frac{2}{m-1}. 
\end{equation} 
The  extrapolated value for $m\rightarrow\infty$ was then found
to be $E_{\infty}= -0.110487$ in good agreement with the DMRG-DMFT
result $E_{\text{DMFT}}(U=4.8)\approx -0.11048$. 
[It is not surprising that we cannot discriminate between the two possibilities~(\ref{eq:scaling1})
and~(\ref{eq:scaling2})
because we actually fit four data points $\{E_m; m=3,5,7,9\}$ using four parameters if we also allow
for the adjustment of the scaling of $x$ with $m$.]
The same analysis was carried out using the 11--th order contribution calculated two years ago~\cite{kal12}
but this additional term alone did not change the results significantly enough to discriminate between 
both fits.

\begin{figure}[tb]
  \includegraphics[width=0.49\textwidth]{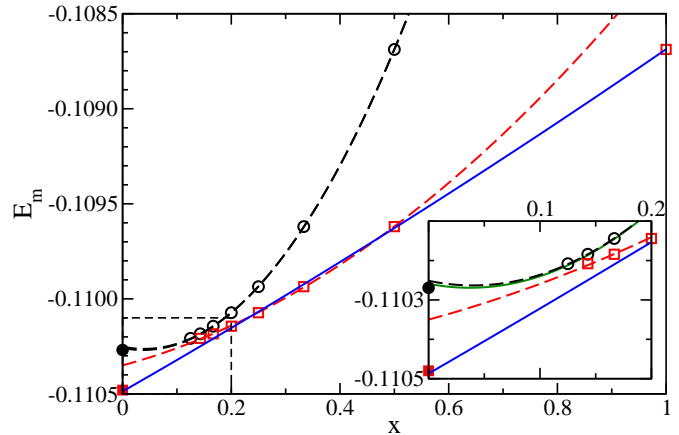}
  \caption{\label{fig2} 
  SCPT ground-state energy $E_m(U=4.8)$ as a function of $x=\frac{2}{m+1}$
  (open circles) and $x=\frac{2}{m-1}$ (open squares). 
  Lines represent least-square quadratic fits~(\ref{eq:fit}) of 
  these data using all points ($m \leq 15$, black or red dashed lines)
  and all but the leftmost two points ($m\leq11$, blue or green solid lines).
  The solid circle and square correspond to the QMC-DMFT and DMRG-DMFT results for $U=4.8$,
  respectively. 
  The inset shows an expanded view of the box in the lower left corner. 
}
\end{figure}

Using the two additional contributions calculated in this work ($m=13$ and $15$) we find that
the fit based on the first scaling~(\ref{eq:scaling1}) remains virtually unchanged from the result
for $m\leq 11$, see Fig.~\ref{fig2}. 
In particular, the extrapolated energy $E_{\infty}= -0.110252$ 
for $U=4.8$ is still in excellent agreement with the QMC-DMFT result.
By contrast, Fig.~\ref{fig2} shows that the fitted parabola based on the second
scaling~(\ref{eq:scaling2}) changes
significantly  if one uses all known data points $E_m$ ($m\leq 15$) or only the previously available ones
($m\leq 11$).
The extrapolated energy $E_{\infty}= -0.110350$ now differs visibly from the DMRG-DMFT result and
shifts closer to the QMC-DMFT result.

We have analyzed the SCPT convergence for various values of $U$ using a more
general scaling $x=2/(m+w)$. 
All results confirm that the choice $w=1$ yields the most stable extrapolations~(\ref{eq:fit}) 
and that the extrapolated SCPT energies $E_{\infty}$ agree very well
with the QMC-DMFT energies for $U \geq 4.8$. 
Moreover, they confirm that the agreement between SCPT and DMRG-DMFT energies
deteriorates for $U\leq 5$ when the orders $n=13$ and $n=15$ are taken into account 
even if one chooses another parameter $w$.
Therefore, we conclude that (\ref{eq:scaling1}) is the best scaling 
for extrapolating ground-state energies and that the QMC-DMFT calculations~\cite{blu05}
are more accurate than the DMRG-DMFT computations~\cite{nis04} in the critical
region above $U_c$.

\subsection{Extrapolated perturbation theory}

Rather than extrapolating the ground-state energy for a given coupling $U$,
we can use the Domb-Sykes method~\cite{dom57,dom61} 
to conjecture the asymptotic behavior of the coefficients $a_n$.
Thus we can obtain the critical behavior of the ground-state energy and also
extrapolate the partial sums~(\ref{eq:partial}) to very high orders $m$.
This approach was named extended perturbation theory (ePT) in previous
works.~\cite{blu05,kal12}

\begin{figure}[tb]
  \includegraphics[width=0.49\textwidth]{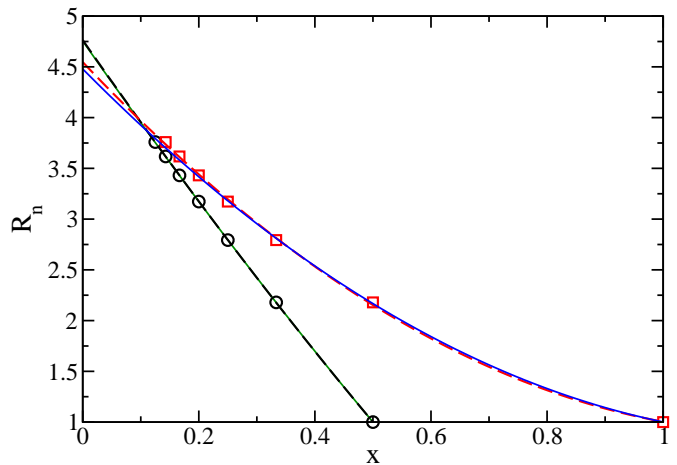}
  \caption{\label{fig3} 
   Domb-Sykes plot of the ratio $R_n$ as a function of $x=\frac{2}{m+1}$
   (open circles) and $x=\frac{2}{m-1}$ (open squares).
   Lines represent quadratic least-square fits~(\ref{eq:fit2}) of these data using all 
   points ($m \leq 15$, black or red dashed lines) and all but the leftmost two points 
   ($m\leq11$, blue or green solid lines).
}
\end{figure}

For odd $n \geq 3$ we define the number sequence
\begin{equation}
\label{eq:ratio}
R_n = \sqrt{\frac{a_n}{a_{n-2}}}.
\end{equation}
Assuming that the convergence radius $U_R$ of the series~(\ref{eq:sum}) 
is identical with the critical coupling $U_c$ where
the Mott phase becomes unstable, the ratio criterion implies that 
\begin{equation}
\lim_{n\rightarrow \infty} R_n = U_c .
\end{equation}
To extrapolate the sequence for $n \rightarrow \infty$,
one can again use a least-square quadratic fit 
\begin{equation}
\label{eq:fit2}
R_n=U_{c} + g_1 x + g_2 x^2
\end{equation}
with $x=2/(n+w)$, and the three fit parameters $U_c,g_1,g_2$.
The corresponding Domb-Sykes plots are shown in Fig.~\ref{fig3}.
If we assume that the singular part of the ground-state energy~(\ref{eq:energy})
for $U \alt U_c$ is a power law
\begin{equation}
E_c(U)  \propto \left ( U-U_c\right )^{\tau-1}
\end{equation}
with a critical exponent $\tau \neq 1,2,3, \dots$, 
the coefficients of the series~(\ref{eq:series})
must satisfy the asymptotic relation 
\begin{equation}
R_n \approx  U_c \left (1 - \frac{\tau}{n} \right )
\end{equation}
for $n\gg 1$.~\cite{dom57,dom61,hun80}
Therefore, we can estimate the critical exponent from the fit parameters
with
\begin{equation}
\tau = -\frac{2 g_1}{U_c} .
\end{equation}

On the basis of the SCPT coefficients $a_n$ up to $n=9$, it was shown
that $U_c\approx 4.75$ and $\tau\approx 3.44$ using $w=1$~\cite{blu05}
but $U_c\approx 4.43$ and $\tau\approx 2.61$ using $w=-1$~\cite{nis04}.
On the one hand, the former critical parameters agreed well with several DMFT calculations for
$U_c$~\cite{bul01,blu02,gar04,blu05,kar05} but none of these studies proposed a value
for $\tau$.
On the other hand, the latter critical parameters were in excellent agreement with 
the DMRG-DMFT results $U_c\approx 4.45$ and $\tau=5/2$.~\cite{nis04}
Moreover, the value $U_c\approx 4.406$ was obtained from a perturbative solution
of the DMFT self-consistency problem.~\cite{ruh11}
Including the 11-th order coefficient did not change the critical parameters
significantly and hence did not solve the controversy.~\cite{kal12} 

Using the two additional orders computed in this work, we find that
the critical parameters are only insignificantly modified,
$U_c\approx 4.76$ and $\tau\approx 3.45$, for the choice $w=1$. 
For $w=-1$, however, the fit parabola becomes visibly different
for $x\rightarrow 0$, see Fig.~\ref{fig3},
and the resulting critical parameters are now $U_c\approx 4.55$ and
$\tau\approx 2.55$.
While the change of $\tau$ is negligible (and the new value agrees rather better with the
DMRG-DMFT $\tau=5/2$ than previously), the critical coupling $U_c$ shifts
significantly away form the DMRG-DMFT result~\cite{nis04} toward the value obtained in 
other DMFT calculations.
We have also probed other values of $w$ but clearly the choice 
$w=1$ yields the most stable extrapolation with respect to variations of the numbers of
exact coefficients $a_n$.
Therefore, we conclude that the critical parameters are $U_c\approx 4.76$ and
$\tau\approx 3.45$ based on the 15-th order SCPT and the Domb-Sykes method.

Assuming that the relation (\ref{eq:fit2}) holds for all coefficients $R_n$ with $n > 15$
we can compute the coefficients $a_n$ for $n>15$ recursively
and
thus extend the partial sum~(\ref{eq:partial}) to very high orders $m$.
Additionally, as we know the asymptotic behavior of the coefficients
\begin{equation}
\label{eq:coeff}
a_n = -U_c^n \frac{C}{n^{\tau}} 
\end{equation}
with a constant $C\approx 0.349$,
we can easily estimate the cutoff $m$ for a given $U$ and accuracy goal.
In Ref.~\onlinecite{blu05}, it was shown using this extrapolated perturbation
series  (and the exact coefficients up to
$n=9$) that the resulting ground-state energies agree with the QMC-DMFT data
within $10^{-5}t$.
Using the additional exact coefficients $a_n$ up to $n=15$ and extrapolated ones up to
$m=1001$, we find that the differences
between the extrapolated perturbation series and the QMC-DMFT ground-state energies 
are  now of the order of $10^{-7}t$ or smaller for all $U-U_c \geq 0.04$.
Therefore, we have not only confirmed that the QMC-DMFT data are numerically exact
(i.e., within their stated precision of $10^{-8}t$) but also that the 
extrapolated perturbation series can reach the same level of accuracy even very
close to the critical coupling.

\section{Conclusion and outlook \label{sec:conclusion}}

We have investigated the ground-state energy in the Mott insulating phase of
the Hubbard model on a Bethe lattice with infinite coordination number
using a combinatorial-diagrammatic approach based on the Kato-Takahashi
strong-coupling perturbation theory.
First, we have carried out large-scale computer-algebra calculations to obtain
the exact coefficients $a_n$ of the series expansion~(\ref{eq:series}) up to the 15-th order in $1/U$.
Then, a Domb-Sykes analysis of the series asymptotic behavior has allowed us
to determine its singular behavior close to the critical coupling $U_c$ below which the Mott phase
becomes unstable. We have thus established highly accurate benchmarks for DMFT methods. 

The DMFT method~\cite{vol11,vol12}  is a complex numerical technique and
the result quality depends not only on
the impurity solver used (e.g., DMRG, QMC or NRG) but also on the chosen discretization scheme
for the continuous self-consistency equation. As DMRG is a very reliable method for
quantum impurity problems and other DMRG-DMFT investigations~\cite{gar04,kar05} agree with QMC-DMFT results,
the failure of the DMRG-DMFT computation close to the critical coupling in Ref.~\onlinecite{nis04}
is probably due to the discretization scheme used in that work.
Indeed, an essential step of this particular scheme is the deconvolution of the
impurity density of states calculated with DMRG.
In a recent work~\cite{pae14}, two of us have shown that the deconvolution procedure used in
 Ref.~\onlinecite{nis04} slightly distorts the shape of the density of states
 in a one-dimensional paramagnetic Mott-Hubbard insulator. We think that 
 a similar deconvolution inaccuracy could be responsible for the failure of the
 DMRG-DMFT scheme in Ref.~\onlinecite{nis04} .
 
The computer-algebra SCPT method presented in Sec.~(\ref{sec:method}) can be extended to various
generalizations of the Hubbard model~(\ref{eq:hamiltonian}).
For instance, one could vary $t_{\downarrow}$ continuously to interpolate
between the Hubbard model ($t_{\downarrow}/t_{\uparrow}=1$) and the Falicov-Kimball model 
($t_{\downarrow}/t_{\uparrow}=0$) or one could study Hubbard models with several bands~\cite{jak09,gre13} or 
internal $SU(n)$ symmetries with large $n$.~\cite{bon12,blu13}
Kato perturbation theory has already been applied to the Mott insulating phase in the strong-coupling limit
of the Bose-Hubbard model with spinless bosons.~\cite{tei09,eck09,hei12}
In that case, the unperturbated ground state is not degenerate and thus the
perturbation series can easily be computed up to high orders. 
For bosons with spin~$S > 0$, however, the degeneracy of the unperturbated ground state 
duplicates the situation encountered in the Hubbard model for electrons. 
Thus, the approach that we have used for the fermionic Hubbard model can also be applied to a 
spin-disordered Mott phase in the Bose-Hubbard model with spin~$S >0$ as well
as to Mott phases of other fermion systems and of boson-fermion mixtures in optical lattices.~\cite{fis89,alt12,blo08}

However, an essential condition for the computer-algebra SCPT used in this work
is the conservation of the singlet ground-state degeneracy at all orders in the series
expansion, which allows one to evaluate the correlation functions~(\ref{eq:correlations}) easily.
If this property is not fulfilled, an exact calculation of the relevant correlation functions
could become much more difficult or even impossible. Then one would have to be content with 
(possibly numerical) approximations for the coefficients $a_n$ of the perturbation series.
Obviously, this degeneracy at finite coupling is a model property and thus the applicability of 
our method has to be checked on a case to case basis.
The loss of degeneracy seems also to be the most serious difficulty in extending the
combinatorial-diagrammatic approach to the Hubbard model away from half filling
and, more generally, to metallic phases in the strong-coupling limit.
Indeed, away from half filling the degeneracy of the $U=\infty$ ground state is
already partially lifted in first order in the hopping term $T$.
Again, this seems to imply that one could only obtain approximate series coefficients $a_n$ away from half filling.
Similarly, we could employ the computer-algebra SCPT  as an approximation method for
the Hubbard model on other lattice geometries than the Bethe lattice and 
for finite dimensions or coordination numbers. (The DMFT method is already used as an approximative 
method for treating strong electronic correlations in finite dimensional systems, for instance,  in first-principles studies of three-dimensional systems.~\cite{vol12,pav11,lin11})

One of the open problems in the theory of Mott insulators is the shape of the Hubbard bands in the
single-particle density of states (DOS). 
In particular, DMFT calculations reveal some unexplained sharp structures at the low-energy edges of 
the Hubbard bands in both the Mott insulating phase~\cite{nis04} and the metallic phase~\cite{kar05,zit09} 
in the critical region. The DOS of the Mott insulating phase has been calculated perturbatively up to the second
order in~$1/U$ directly from the Hubbard model~\cite{eas03} and up to the third order by solving the DMFT 
self-consistency equation~\cite{ruh11}. However, these results do not fully explain the observed structures. 
Moreover, the DMFT results for the DOS depend sensitively on the scheme used to solve the self-consistent 
impurity problem~\cite{kar05}. 
In that case we are clearly in need of more accurate results, such as
higher-order terms in the perturbation expansion.

In principle, the combinatorial-diagrammatic approach can be extended to the calculation of the local single-particle
Green's function, which determines the DOS and the Mott-Hubbard gap.
The series expansion can be formulated as a
self-consistent integral equation for the Green's function at finite $U$.
The equation contains polynomials of the Green's function at $U \rightarrow
\infty$ with increasing orders.
The coefficients of these polynomials can be calculated using a similar
combinatorial-diagrammatic approach as the coefficients for the series expansion of the ground-state 
energy~(\ref{eq:series}).  
However, the computational cost appears to be significantly higher for the
Green's function than for the ground state series expansion.
Moreover, it is not clear whether we can obtain an exact solution with
combinatorial-diagrammatic techniques only,
because methods from numerical analysis could be required to solve the self-consistent integral equation. 
Nevertheless, it would be worthwhile to calculate even only a few higher order contributions to the Green's function.
Knowing
 higher-order contributions to DOS and gap would allow us to determine the critical coupling $U_c$ and the
critical exponent $\tau$ more accurately and thus to gain a better understanding of the paramagnetic Mott metal-insulator transition.
Moreover, this would provide us with a more direct and thorough benchmarking of numerical DMFT methods
because they are actually based on self-consistent computations of the Green's function.

The development of the  combinatorial-diagrammatic approach to the Kato-Takahashi SCPT has greatly benefited
from a formal mathematical study of its algorithms.~\cite{loogen,gru11}
Discrete mathematics rather than differential calculus provides the mathematical background for 
this approach.
Further development of similar computer-algebra perturbation methods will require 
a close cooperation between  physics and discrete mathematics which will benefit both fields.
Indeed, we have not only used the On-Line Encyclopedia of Integer Sequences (OEIS)~\cite{oeis}
to obtain information on known integer sequences but also contributed new ones.
For instance, the number of sequences $N_s$ in Table~\ref{table1} is the integer sequence A198761 
in OEIS.

The computer-algebra techniques developed in this work for large-scale computations of the Kato-Takahashi SCPT
could also be applied to other series expansions.~\cite{oit06,apel,hag07,che04} For instance, 
the method of continuous unitary transformations can be used to map the Hubbard model at strong coupling
onto an effective model with conservation of the number of double occupancies~\cite{ste97,rei04,ham10}. 
Using appropriate truncation schemes one can close, and thus solve, the flow equations~\cite{keh06} of the 
effective Hamiltonians. 
This results in a systematic expansion of the effective Hamiltonian and other observables in powers of~$1/U$, 
which is very similar to  Kato perturbation expansion.
One possible approach is a truncation of the equations in a perturbative manner to obtain a
series expansion.~\cite{yan10} Recently, a non-perturbative approach has been proposed
based on graph-theoretical methods.~\cite{yan11}
Therefore, we think that larger-scale computer-algebra calculations will also prove useful for
these approaches in the future.

\begin{acknowledgments}

We are very thankful to R.~Loogen and G.~Gruber for 
examining the combinatorial--diagrammatic algorithm 
and for numerous helpful discussions on graph theory and combinatorics. 
We thank N.~Sloane and A.~Heinz for pointing out the number
theory aspect of the combinatorial-diagrammatic method as well as
F. Gebhard and K. Schmidt for useful discussions regarding the Kato-Takahashi perturbation theory. 
We are indebted to G.~Gaus, G.~Brand, and M.~Brehm for their assistance with the porting and testing of our program.
Computer resources for this work were provided by the Hewlett-Packard Development Company, L.P., the North-German
Supercomputing Alliance (HLRN), and the Leibniz Supercomputing Centre (LRZ).

\end{acknowledgments}

\end{document}